\documentclass[manuscript,nonacm]{acmart}
\renewcommand\footnotetextcopyrightpermission[1]{}

\usepackage{todonotes}
\usepackage{multirow}
\usepackage{booktabs}
\usepackage{mathtools}
\usepackage{enumitem}
\usepackage[table]{colortbl}
\usepackage{cleveref}
\usepackage{framed}
\usepackage{xspace}
\usepackage{graphicx}
\usepackage{caption}
\usepackage{subcaption}
\usepackage{lipsum}
\usepackage{soul}
\usepackage{float}
\usepackage{makecell}
\usepackage[table]{xcolor}
\usepackage{caption}

\begin{document}

\title[Understanding Users' Privacy Reasoning and Behaviors During Chatbot Use]{Understanding Users' Privacy Reasoning and Behaviors During Chatbot Use to Support Meaningful Agency in Privacy}

\author{Mohammad Hadi Nezhad}
\email{mhadinezhad@cs.umass.edu}
\orcid{0009-0002-2750-8296}
\affiliation{
  \institution{University of Massachusetts Amherst}
  \city{Amherst}
  \state{MA}
  \country{USA}
}

\author{Francisco Enrique Vicente Castro}
\orcid{0000-0002-3940-5334}
\affiliation{
  \institution{New York University}
  \city{New York}
  \state{NY}
  \country{USA}
}

\author{Ivon Arroyo}
\orcid{0000-0002-9697-8016}
\affiliation{
  \institution{University of Massachusetts Amherst}
  \city{Amherst}
  \state{MA}
  \country{USA}
}

\renewcommand{\shortauthors}{Hadi Nezhad et al.}

\begin{abstract}
Conversational agents (CAs) (e.g., chatbots) are increasingly used in settings where users disclose sensitive information, raising significant privacy concerns. Because privacy judgments are highly contextual, supporting users to engage in \emph{privacy-protective actions} during chatbot interactions is essential. However, enabling meaningful engagement requires a deeper understanding of how users currently reason about and manage sensitive information during realistic chatbot use scenarios. To investigate this, we qualitatively examined computer science (undergraduate and masters) students' in-the-moment disclosure and protection behaviors, as well as the reasoning underlying these behaviors, across a range of realistic chatbot tasks. Participants used a simulated ChatGPT interface \textit{with} and \textit{without} a \emph{privacy notice panel} that intercepts message submissions, highlights potentially sensitive information, and offers privacy-protective actions. The panel supports anonymization through retracting, faking, and generalizing, and surfaces two of ChatGPT's built-in privacy controls to improve their discoverability. Drawing on interaction logs, think-alouds, and survey responses, we analyzed how the panel fostered privacy awareness, encouraged protective actions, and supported context-specific reasoning about what information to protect and how. We further discuss design opportunities for tools that provide users greater and more meaningful agency in protecting sensitive information during CA interactions.

\end{abstract}


\maketitle

\section{Introduction}
People engage with Conversational Agents (hereafter CAs or chatbots) such as ChatGPT across a range of domains, including professional work, academic spaces, finance, healthcare, and casual conversations \cite{NBER-2025,zhang-2024,zhang-2025-exploring,esmaeili-2024}. 
At the same time, scholars have emphasized growing public concerns that the expansion of AI systems (including CAs) could erode privacy, particularly due to extensive collection of highly personal data that enables detailed profiling and creates risks of data misuse beyond users' consent \cite{kelley-2023,mutahar-2025}. 
In response, research has focused on privacy protection \textit{after} users interact with CAs and across stages of data processing \cite{yu-2024,kandpal-2022,ait-mlouk-2023,jang-2023-knowledge,majmudar-2022,bagdasarian-2024}, yet users' control over these processes is typically limited, with control options often implemented with dark UX designs that hinder discoverability and effective use \cite{zhang-2024,malki2025hoovered,mutahar-2025-understanding}.
Other research has approached privacy \textit{during} CA interactions by making users aware of privacy (e.g., the presence of sensitive information in their prompts, risks of privacy breaches) and supporting them in taking protective actions \cite{Zhou-2025-rescriber,Chen-2025-CLEAR,Chong-2025-casper,zhang2024adanonymizer}. 
Recognizing that privacy protection is inherently contextual and subjective \cite{nissenbaum-2004,schaar-2010}, user-facing approaches support users in reasoning about tools' privacy policies, context of use, and personal privacy preferences to engage in actions that reduce unnecessary data leaving their devices during CA interactions.
However, supporting users to exercise meaningful agency over their privacy requires a deeper understanding of how they currently behave and think about disclosing or protecting sensitive information as interactions unfold, especially under realistic use scenarios (e.g., when users are not primed toward studies' privacy goals, when activities like prompting and evaluating responses impose high cognitive demands \cite{lev-2024}).
Such insights can equip CA designers in addressing users' existing privacy concerns and needs, ultimately fostering greater and meaningful engagement in privacy-protective actions. 

Motivated by this need, we build on prior work by qualitatively investigating in-the-moment sensitive information disclosure and protection behaviors of computer science (CS) (undergraduate and masters) students, as well as their real-time reasoning behind these behaviors.
We focus on students as they are frequent chatbot users \cite{zhang-2025-exploring,stohr-2024}, younger users are more prone to disclosing sensitive information \cite{lappeman-2022-trust}, and CS students can---as potential CA developers---play a critical role in shaping privacy-conscious AI systems \cite{hadinezhad-2025}.
We conducted two sessions wherein participants completed tasks: one to examine baseline behaviors when using a simulated \textit{ChatGPT} interface, and another to examine how these behaviors and reasoning might change when the simulation is integrated with a just-in-time \emph{privacy notice panel}.
Our panel intercepts interactions right after an attempt to send a message containing sensitive details, encouraging context-based decision-making.
It identifies all instances of certain sensitive information embedded in the task materials (e.g., \textit{names},\textit{ physical addresses}) while ignoring others (e.g., health or financial details), enabling us to observe whether participants take protective steps beyond what the panel provides.
The panel also offers multiple protection options, including three anonymization strategies for flagged items (\textit{retracting}, \textit{faking}, \textit{generalizing}), and two of the \textit{ChatGPT}'s built-in privacy controls (\textit{opting in/out of sharing content for model training} and \textit{enabling/disabling memory})---surfaced through the panel to improve discoverability \cite{Brown-2022-whatdoes,zhang-2024}. 
This allowed us to analyze behaviors and reasoning across a wide range of approaches. 
Finally, our panel supports carrying out anonymization strategies at two levels of granularity---per instance or across all instances of a given information type, aiming at examining how deeply participants contextualize their choices. 
Participants then completed a post-study survey where they reflected on their tasks during the sessions.

To closely approximate realistic chatbot use, our study: (1) incorporates varied real-world use scenarios (e.g., writing emails, searching lengthy documents, summarizing text) through user tasks involving embedded data with different types of sensitive information; (2) withholds the study's privacy goals from participants until after participation to reduce biases; and (3) allows participants to freely direct their interactions (e.g., prompting, evaluating responses, disclosure) under approximated time and effort constraints (e.g., planning a one-hour session). 
Thus, we examine four RQs regarding real-time interactions with CAs:
\vspace{0.5em}
\begin{enumerate}[label=\textbf{RQ\arabic*.}, leftmargin=3em]
    \item How do students disclose or withhold sensitive information without vs. with the privacy notice panel?
    \item Why do students disclose or withhold sensitive information when interacting with the chatbot?
    \item How do students use privacy-protective approaches without vs. with the privacy notice panel? 
    \item Why do students use or avoid specific privacy-protective approaches when interacting with the chatbot?
\end{enumerate}
We collected data from ten students, including audio and video recordings of the sessions, think-aloud transcripts, logs of chatbot interactions, and survey responses. 
Through a detailed qualitative analysis (Section \ref{subsec:analysis}), we present common behaviors regarding disclosure of sensitive information (\textit{RQ1}; Section \ref{sec:rq1}), use of protective approaches (\textit{RQ3}; Section \ref{sec:rq3}), and students' common rationale underlying these behaviors (\textit{RQ2}; Section \ref{sec:rq2}, \textit{RQ4}; Section \ref{sec:rq4}).
We observed that, without the panel, interactions were largely task-focused with limited concern for privacy. 
In contrast, the panel promoted privacy awareness and protective behaviors, either through the panel or manual efforts. 
Across these behaviors, participants considered a range of contextual factors for reasoning about what to protect and how.
We discuss opportunities to further enhance meaningful user engagement and support in preserving privacy during CA use (Section \ref{sec:discussion}). 
These opportunities target key phases of users’ decision-making, including supporting privacy awareness and reasoning, supporting the filtering of sensitive information that is irrelevant to task goals, balancing automation with user control in anonymization, and encouraging manual protections when privacy tools are limited.

\section{Background and Related Work}
\label{sec:rw}
\subsection{Sensitive Information Disclosure and Privacy Risks in Conversational Agents (CAs)}
\label{subsec:rw_info_disclosure}
In interactions with CAs, users frequently disclose sensitive information about themselves or others, including personally identifiable information (PII) (e.g., names, email addresses), less directly identifiable information (e.g., nationality, religious affiliation), and details about activities or behaviors \cite{zhang-2024,mireshghallah-2024,malki2025hoovered}. 
Such disclosures can occur in tasks such as seeking medical or exercise advice, drafting emails that contain personal data, or sharing source code with embedded confidential information to fix errors.
As users share more data with CAs, they have become major repositories of personal information, yet they may not be adequately equipped to minimize privacy risks. Large Language Model (LLM)-based CAs are especially vulnerable to the \emph{memorization effect}, where sensitive training data can be reproduced through targeted prompts \cite{nasr-2025,carlini-2023,Brown-2022-whatdoes}. 
Studies also highlight broader privacy risks introduced or exacerbated by AI systems, such as risks related to data collection (e.g., surveillance), processing (e.g., identification, aggregation), and dissemination (e.g., exposure, distortion) \cite{lee-2024}. 
Well-documented incidents further illustrate these risks. 
For example, the Korean chatbot \textit{Lee Luda} was accused of leaking personal details (e.g., addresses, bank accounts) from the \textit{KakaoTalk} data used in its training \cite{AIAAIC-LeeLuda}.
In 2023, reports revealed that private conversations with \textit{Bard} (now \textit{Gemini}) appeared in Google search results, later linked to the chatbot's “conversation share” functionality \cite{AIAAIC-bard}.

\subsection{Supporting Users in Protecting Sensitive Information \textit{During} Interactions with CAs}
\label{subsec:rw_supporting_users}
\textit{Contextual integrity} \cite{nissenbaum-2004} indicates that privacy is preserved by two informational norms: \textit{distribution} and \textit{appropriateness}. 
Norms of \textit{distribution} govern the flow of information from one party to another and primarily relate to protections applied after collecting information. 
In CAs, researchers have explored post-collection safeguards during model training (e.g., data sanitization) \cite{yu-2024,kandpal-2022}, after training (e.g., knowledge unlearning) \cite{ait-mlouk-2023,jang-2023-knowledge}, and during inference (e.g., differentially private decoding) \cite{majmudar-2022,bagdasarian-2024}. 
Yet users typically have limited and often ineffective control over these processes, partly due to dark UX designs. 
For instance, in \textit{ChatGPT}, user content is used by default for model training; although an opt-out setting exists, it is shown to be difficult to discover or use effectively \cite{zhang-2024,malki2025hoovered,mutahar-2025-understanding}. Our findings show that surfacing the opt-out feature by the privacy panel enhanced participants' awareness and use of this feature (Sections \ref{sec:rq3_without} and \ref{results_builtin_controls}).

Despite their importance, post-interaction protections have clear limitations. Once data leave users’ devices, developers struggle to accommodate individual privacy preferences (e.g., differences in what individuals consider sensitive \cite{naeini2017privacy}) and may need significant resources to identify and protect such information. 
Users must also trust multiple parties handling their data, as evident in users' concerns about human reviewers accessing conversations, third-party access, and limited transparency and control over data usage \cite{mutahar-2025}.
Related to these, norms of \textit{appropriateness} concerns what information are suitable to disclose in a given context. Similarly, the \textit{Privacy by Design} approach \cite{schaar-2010} stresses minimizing personal data collection and use. 
These norms and approaches highlight that privacy protection is contextual, requiring ongoing and meaningful user involvement in preserving privacy.
However, users’ disclosure of sensitive information to CAs (i.e., self-disclosure) can be intensified by several factors, often without users’ conscious awareness or careful consideration of privacy risks.
These factors include perceived well-being benefits of self-disclosing (e.g., reducing stress and anxiety) \cite{joinson-2001,reis-2017}; interactional cues that encourage disclosure, such as a sense of anonymity, a non-judgmental interaction style, and an appearance of empathy and accuracy in responses \cite{ho-annabell-2018,croes-2024}; anthropomorphic design of chatbots that can increase trust and disclosure \cite{2024-maeda,gabriel2024ethics,2020-Ischen}; and the convenience of sharing lengthy documents \cite{zhang-2024}.
To mitigate these influences, researchers have developed interfaces that promote users' privacy awareness and protective actions during CA interactions. 
For example, \textit{Rescriber} \cite{Zhou-2025-rescriber} identifies personal information in messages and offers \textit{redaction} or \textit{abstraction} prior to submission. 
\textit{CLEAR} \cite{Chen-2025-CLEAR} detects sensitive information, summarizes relevant privacy policies, and uses LLMs to present potential disclosure risks. 
\textit{Casper} \cite{Chong-2025-casper} automatically replaces PIIs with placeholders and warns users about sensitive topics in their prompts. 
Building on these efforts, our just-in-time privacy panel offers three anonymization strategies (\textit{retracting}, \textit{faking}, \textit{generalizing}), applicable at two levels of granularity (per instance or across all instances of an information type), allowing us to examine how users employ these strategies across varied tasks.

\subsection{Users’ Disclosure and Protection Behaviors and Their Reasoning}
\label{subsec:rw_users_behaviors}
Zhang et al. \cite{zhang-2024} analyzed histories of \textit{ChatGPT} conversations and identified types of sensitive information users commonly disclosed and the protective strategies evident in their chats.
Through interviews with \textit{ChatGPT} users, they further identified factors shaping disclosure decisions (e.g., perceived CA task capability, disclosure risks and harms, disclosure norms, and flawed mental models of how CAs process data), as well as how users navigate trade-offs between disclosure risks and benefits (e.g., accepting privacy risks to reap benefits, manually sanitizing data, avoiding tasks).
Overall, they found that users continually consider trade-offs between privacy, chatbot utility, and convenience of protecting privacy (see also \cite{Brown-2022-whatdoes}). 
In the \textit{Rescriber} study \cite{Zhou-2025-rescriber}, participants completed interview-based sessions in which they used \textit{ChatGPT}, with \textit{Rescriber} operating in the browser, to work through one researcher-created scenario and one participant-prepared conversation involving sensitive data. 
The authors observed that users balanced information sensitivity with task relevance when deciding what to sanitize. 
For protection actions, participants tended to \textit{abstract} sensitive information when context was needed and use \textit{redaction} for non-essential sensitive details.
Building on prior work, we examine participants' \textit{in-the-moment} disclosure and protection behaviors and the reasoning that guides these decisions as they complete tasks under closely approximated real-world chatbot use scenarios.

\section{Methods}
\label{sec:methods}
In this section, we describe our \textit{ChatGPT} interface simulation and privacy notice panel design (Section \ref{sec:system_design}), the user study design (Section \ref{subsec:study_design}), participant recruitment and data collection procedures (Section \ref{subsec:participants_data}), and our analysis approach (\Cref{subsec:analysis}).
All research procedures described here were reviewed and approved by our university IRB. 

\subsection{ChatGPT Interface Simulation and Privacy Notice Panel Design}
\label{sec:system_design}
We built an interactive simulation of \textit{ChatGPT}'s interface (as of Fall 2024) connected to the \textit{GPT-4o} API. 
The simulation includes \textit{ChatGPT}'s core interface components including a message input box, a conversation panel, and static icons on the top right and left (Appendix \ref{appendix_interface_without}, Figure \ref{fig:without_panel}). An interactive \textit{profile icon} on the top right opens a settings panel (mirroring \textit{ChatGPT}'s design) with two working tabs: (1) \emph{Data Control}, for opting in/out of sharing content for model training, and (2) \emph{Personalization}, for enabling or disabling chatbot’s memory (Appendix \ref{appendix_interface_without}, Figure \ref{fig:settings-panel}).
By default, the memory feature retains the last seven messages; when disabled, it considers only the most recent message. 
While not replicating \textit{ChatGPT}’s actual memory system, this design exposes participants to tradeoffs between utility (retaining context) and protecting privacy (disabling memory). 
We developed a privacy notice panel and integrated it into our simulation.
The panel intercepts user messages immediately after submission attempts (i.e., the submit button is clicked) and prior to transmission to the \textit{ChatGPT}'s API. 
The panel appears on the right side of the screen, without blocking the ongoing interaction, and its use remains optional. 
The detection engine operates locally within the user’s browser and is capable of recognizing all \textit{names of people}, \textit{email addresses}, \textit{phone numbers}, \textit{physical addresses}, \textit{social security numbers}, and \textit{dates of birth} embedded in our user tasks (Section \ref{subsubsec:user_tasks} and Table \ref{tab:task-overview} in Appendix \ref{appendix_user_tasks}).

\begin{figure}[h]
  \centering
  \includegraphics[width=\linewidth]{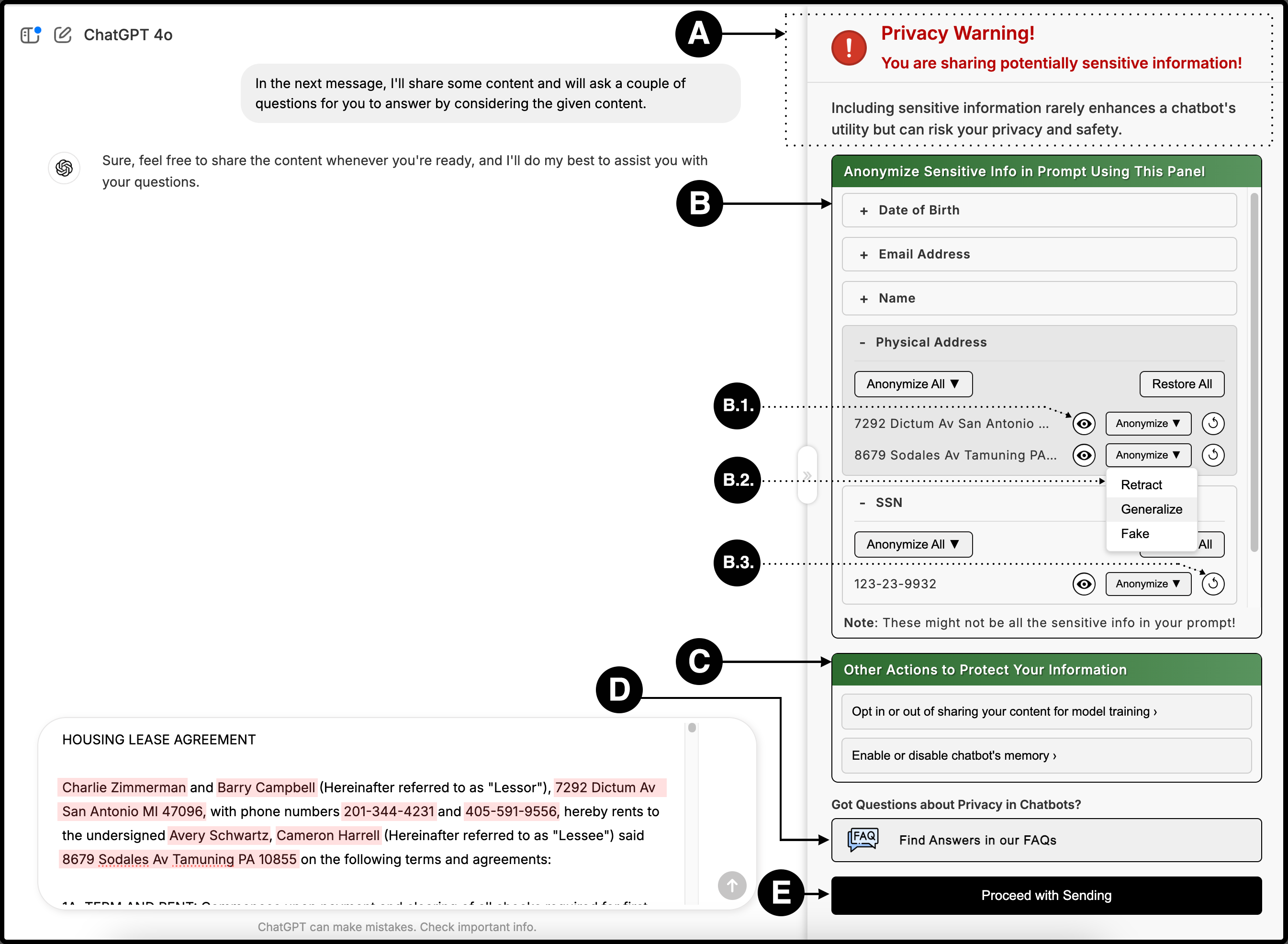}
  \caption{ChatGPT Interface Simulation \textit{with} Privacy Notice Panel. The panel appears after an attempt to send a message containing sensitive information (highlighted in the input box). It includes (A) a warning message, (B) an anonymization panel, (C) shortcuts to built-in privacy controls, (D) FAQs about privacy in CAs, and (E) a proceed with sending button. The anonymization panel further includes, for each detected instance, a (B.1.) locate icon, (B.2.) drop-down menu of anonymization options, and (B.3.) restore button.}
  \label{fig:with_panel}
  \Description{Screenshot of the simulated ChatGPT interface showing the integrated privacy notice panel. The panel appears after a user attempts to send a message containing sensitive information (the sensitive information identified in a sample message is highlighted). The panel is full-height, positioned on the right side of the screen, next to the input box and conversation panel. From top to bottom, the panel includes: (A) a warning section with the heading ‘Privacy Warning!’ and messages: ‘You are sharing potentially sensitive information’ and ‘Including sensitive information rarely enhances a chatbot's utility but can risk your privacy and safety’; (B) an anonymization panel titled ‘Anonymize sensitive info in prompts using this panel’, listing detected types of sensitive information (e.g., date of birth, physical address). Two entries (i.e., types of info) are expanded to show features: (B.1) a locate icon, (B.2) a dropdown menu of anonymization options (retract, generalize, fake), and (B.3) a restore icon. A note at the bottom states: ‘These might not be all the sensitive info in your prompt!’; (C) a section titled ‘Other actions to protect your information’ and includes shortcut buttons for two built-in privacy controls named as ‘opt in or out of sharing content for model training’ and ‘enabling or disabling chatbot memory’; (D) a section with the heading ‘Got questions about privacy in chatbots?’ followed by a button labeled ‘Find answers in our FAQs’; and (E) a button to proceed with sending.}
\end{figure}

The privacy panel displays (annotated correspondingly in \Cref{fig:with_panel}):
(A) \textit{Warning Message}--notifying the user of potentially sensitive information in their message, (B) \textit{Anonymization Panel}--listing the types of sensitive information identified in a message, (C) \textit{Shortcuts to ChatGPT's Built-in Privacy Controls}, (D) \textit{FAQs}--Q/As about privacy in CAs; we do not discuss this further as it is beyond the study's scope, and (E) \textit{Proceed with Sending}--sending the current message to the \textit{ChatGPT}'s API. 
Each type of sensitive information listed in the \textit{Anonymization Panel} (B) is expandable, showing the detected instances. For each instance, users can: (B.1.) \textit{Locate and Highlight} it in their message, (B.2.) Open a \textit{Drop-down Menu} to anonymize the instance, and (B.3) \textit{Restore} the original value. 
There are three anonymization options (B.2): \textit{Retracting}--replacing with a type label (e.g., \textit{341-23-9932} $\rightarrow$ \textit{[SSN]}), \textit{Generalizing}--retaining \textit{US state} and \textit{ZIP} for addresses, \textit{year} for dates of birth, and \textit{Faking}--replacing with a dummy value (e.g., actual name \textit{Cameron} $\rightarrow$ fake name \textit{Arron}).
Additional controls allow \textit{Anonymizing All} and \textit{Restoring All} instances of a given type at once. A note at the bottom of the \textit{Anonymization Panel} (B) reminds users that the list may not capture all sensitive information in their message.

\subsection{Study Design}
\label{subsec:study_design}
Each participant completed three sequential phases.
We adopted an \textit{incomplete disclosure} strategy (i.e., withholding the study’s specific privacy goals until after participation) to reduce potential biases. 
Below, we describe each phase of the study and details about the user tasks.

\subsubsection{Study Phases}
In a remote \textit{Zoom}\footnote{https://www.zoom.com/} session, participants completed the first task assignment (Section \ref{subsubsec:user_tasks}) using the chatbot simulation \textit{\textbf{without}} the privacy notice panel, while sharing their screen and thinking aloud. 
The first author facilitated all remote sessions, beginning with an introduction to the study (without mentioning its privacy-related goals), tasks, and the chatbot. 
Participants then received think-aloud instructions and completed a brief practice task. 
The researcher then muted himself and disabled his video to minimize distractions.
The session's duration (1 hour) was intended to balance avoiding participant fatigue while providing sufficient time for varied interaction styles (e.g., anonymizing information, evaluating chatbot responses).
After a short break of 1-4 days (depending on participants' availability), participants completed the alternate task assignment in a similar remote \textit{Zoom} session, using the chatbot simulation \textit{\textbf{with}} the integrated privacy notice panel (Figure \ref{fig:with_panel}).
Within one day, participants completed a post-study survey, capturing their reflections on the task sessions. 
The survey included open-ended questions about chatbot features they found supportive in protecting sensitive information and their thoughts on components of the panel (e.g., the anonymization component).
They were also invited to share any additional comments.

\subsubsection{User Tasks}
\label{subsubsec:user_tasks}
We created two versions of task assignment (A and B) and alternated them across the two task-completion sessions for each participant to counterbalance order and content effects. For example, participants who completed \textit{Assignment A} in the first session completed \textit{Assignment B} in the second session, and vice versa (Table \ref{tab:participants}).
Both versions include comparable tasks involving summarizing text, classifying text, searching within lengthy content, and drafting emails. 
Each assignment consists of three tasks, each with two steps that provide descriptions, necessary data, and a text entry field for submitting responses. 
We designed tasks that: (1) approximate real-world chatbot use scenarios and tailored them to CS undergrads and masters to maintain relevance; (2) place participants in scenarios aligned with common disclosure factors \cite{zhang-2024}---e.g., unintentional disclosure (we embedded sensitive details within lengthy content); and (3) avoid emphasizing privacy to minimize bias (consistent with our \textit{incomplete disclosure} strategy) while using subtle prompts encouraging participants to treat the data as real or as their own. 
A detailed description of our tasks and the embedded sensitive information---types, frequencies, and subjects (who the information pertains or belongs to)---are provided in Appendix \ref{appendix_user_tasks}, Table \ref{tab:task-overview}.
The post-study survey and task assignments were administered through \textit{Qualtrics}\footnote{https://www.qualtrics.com/}.

\subsection{Participants and Data Collection}
\label{subsec:participants_data}
We selected CS undergraduate and master's students as participants because students are frequent users of chatbots (particularly \textit{ChatGPT}) \cite{zhang-2025-exploring,stohr-2024}, younger users are more prone to disclosing sensitive information \cite{lappeman-2022-trust}, and CS students are likely future chatbot developers of chatbots, making it valuable to examine their perspective and improve their understanding of privacy in design;
Limiting the participant population further enabled us to design tasks, surveys, and interface features that aligned with participants’ backgrounds and realistic use scenarios (e.g., one task regarding semester schedules), strengthening the ecological validity of the study.
We recruited participants through a poster distributed at a large public university in the northeastern United States. 
Interested students completed a screening survey that collected background information (e.g., degree level, gender, and \textit{ChatGPT} use). 
Recruited students ranged from third-year undergraduate to second-year master’s students. 
Eleven students were recruited; one completed a pilot session that informed refinements to the study protocol. 
The final sample included ten participants who completed all phases between March and June 2025. 
We paused recruitment after ten participants as our iterative analysis suggested thematic saturation, with no substantively new behaviors or rationales observable in additional sessions.
Each participant received a \$70 USD gift card. Participant details are summarized in \Cref{tab:participants}.
For each participant, we collected (1) responses to both task assignments; (2) responses to post-study survey; (3) audio and video recordings of the two task sessions, including think-aloud transcripts; and (4) logs of chatbot interactions (user prompts and model responses). Each study session---excluding the introduction and practice task---ranged between 15 and 80 minutes.

\begin{table*}
  \caption{Participant Details (self-reported) and their Task Assignment Versions. \textbf{Session 1} refers to the task-completion session using the chatbot \textbf{without} the panel, and \textbf{Session 2} refers to the session using the chatbot \textbf{with} the panel.}
  \label{tab:participants}
  \begin{tabular}{lllllll}
    \toprule
    \textbf{ID} 
    & \textbf{Degree - Year} 
    & \textbf{Gender} 
    & \multicolumn{2}{l}{\textbf{ChatGPT Use}} 
    & \multicolumn{2}{l}{\textbf{Task Assignment Version}} 
    \\
    \cmidrule(lr){4-5}
    \cmidrule(lr){6-7}
    & & & \textbf{Duration} & \textbf{Frequency} & \textbf{Session 1} & \textbf{Session 2} \\
    \toprule    
    P1 & Master - 1st year & Female & ~\textasciitilde2.5 year & Daily & A & B \\
    P2 & Undergrad - 4th year & Female & ~\textasciitilde2.5 years & Daily & B & A \\
    P3 & Master - 1st year & Female & ~\textasciitilde2.5 years & Several times a week & A & B \\
    P4 & Master - 2nd year & Male & ~\textasciitilde2.5 years & Daily & B & A \\
    P5 & Undergrad - 3rd year & Male & ~\textasciitilde1.5 years & Daily & A & B \\
    P6 & Undergrad - 4th year & Male & ~\textasciitilde2 years & Several times a week & B & A \\
    P7 & Undergrad - 4th year & Female & ~\textasciitilde2 years & Several times a week & A & B \\
    P8 & Undergrad - 3rd year & Male & ~\textasciitilde1 year & Once a week & B & A \\
    P9 & Master - 1st year & Male & ~\textasciitilde3 years & Daily & A & B \\
    P10 & Undergrad - 3rd year & Female & ~\textasciitilde1.5 years & Daily & B & A \\
    \bottomrule
  \end{tabular}
\end{table*}

\subsection{Analysis Approach}
\label{subsec:analysis}
We developed a coding template that listed each task along with the types and frequencies of sensitive information it contained, and fields for recording whether such information was disclosed, withheld, or attempted to be disclosed (i.e., attempted submission and intercepted by the panel), as well as any anonymization details (e.g., approaches used and whether by the panel).
The template also included fields for capturing participants' qualitative reasoning---drawn from their utterances and interface interactions---regarding disclosure or withholding decisions and their use or avoidance of specific protective approaches. 
This framework enabled systematic analysis of behaviors and reasoning across factors including information type (e.g., \textit{names}, \textit{SSNs}) and protective approaches (e.g., \textit{retracting}, \textit{disabling memory}).
For each participant, we created separate sheets for each of the task-completion sessions. 
The first author populated the coding sheets by reviewing video recordings, think-aloud transcripts, and chatbot interaction logs, triangulating across data sources to resolve ambiguities (e.g., linking interface actions in videos with corresponding think-aloud explanations). 
We analyzed sheets across and within participants to identify common disclosure behaviors (\textit{RQ1}, Section \ref{sec:rq1}) and use of privacy-protective approaches (\textit{RQ3}, Section \ref{sec:rq3}).
We then analyzed the qualitative data recorded in these sheets along with student responses to the \textit{post-study survey} using inductive thematic coding \cite{braun-2006}, guided by our RQs. 
While our coding process allowed us to identify themes directly from the data, our analysis was oriented toward understanding participants’ rationale for disclosing or withholding (\textit{RQ2}, \Cref{sec:rq2}) and for using or avoiding specific privacy-protective approaches (\textit{RQ4}, \Cref{sec:rq4}).
Throughout the coding processes, the first and second authors held weekly collaborative interpretation meetings to refine the coding and resolve disagreements on ambiguous cases by clarifying each others' interpretations of the data and creating themes that captured shared understanding of the data. 

\section{Results}
\label{sec:results}

\subsection{Behaviors in Disclosing Sensitive Information Across Task-Completion Sessions (RQ1)}
\label{sec:rq1}

\subsubsection{Without Panel: Task-Focused Interactions Without Protecting Sensitive Information}
\label{sec:rq1_without}
When using the chatbot \textit{without} the privacy panel, eight participants (\emph{P2–P7, P9, P10}) showed no evidence of considering sensitive information protection; their interactions were task-focused, and they shared most of the sensitive information in their task material.
Only \emph{P1} aimed to protect sensitive information by occasionally withholding task-irrelevant data likely to contain sensitive details---e.g., omitting the “from” and “to” fields in email data.
While their efforts reduced some disclosures, other sensitive details were still shared, and in later tasks, \emph{P1} no longer withheld any data. 
\emph{P8} used the chatbot only to clarify a few terms and did not engage with it for any core task steps, thereby sharing no sensitive information. 

\subsubsection{With Panel: Increased Awareness of Sensitive Information and Emergence of Manual Sanitization}
When using the chatbot \textit{with} the panel, all participants encountered the panel, making them aware of sensitive details in their messages. 
Five participants (\emph{P3, P4, P6, P9, P10}) attempted to share (i.e., attempted submission, but were intercepted by the panel) \textit{all} sensitive information included in their task assignment, while four (\emph{P1, P2, P5, P7}) manually excluded \textit{some} sensitive details before submitting messages, typically those that were easy to spot (e.g., at the top or bottom of a document). This manual exclusion behavior emerged during the session, after these participants had encountered the panel at least once.

\subsubsection{With Panel: Anonymization of Flagged Sensitive Information}
Of the nine participants who attempted to submit sensitive information (all except \emph{P8}), three (\emph{P1, P3, P7}) used the panel to \textit{anonymize all} flagged items before submission; \emph{P10} \textit{submitted all} flagged items as-is; and five (\emph{P2, P4, P5, P6, P9}) used the panel to \textit{selectively anonymize} some information while sharing others as-is.
\emph{P8} had very limited interaction with the chatbot---only for classifying two emails by sentiments
---and did not use the panel on those occasions, sharing flagged information.

\subsubsection{With Panel: Limited Effort to Protect Unflagged Sensitive Information}
\label{results_unflagged_info_behavior}
Only \emph{P1} proactively looked for and withheld sensitive information beyond what the panel was able to detect, noting: \emph{``I'm just trying to remove the things that I think I won't be prompted to auto [anonymize].''}
Although four other participants (\emph{P2, P3, P7, P9}) noticed such sensitive details (e.g., bank account number, insurance ID) or expressed awareness of possible presence of additional sensitive data, none made any effort to identify or protect them, suggesting a reliance on technological features for privacy protection.

\subsection{Rationale Behind Disclosing or Withholding Sensitive Information (RQ2)}
\label{sec:rq2}
To understand the reasoning behind students' disclosure behaviors (Section \ref{sec:rq1}), we analyzed their in-the-moment verbalized thoughts and responses to the post-study survey. 
Below, we summarize the recurring themes and provide example quotes with task identifiers (e.g., \textit{[A-T2]} indicates the second task in \textit{Assignment A}, described in Appendix \ref{appendix_user_tasks}).

\subsubsection{Relevance of Sensitive Information to the Task (P1, P2, P4, P5, P6, P7, P9, P10).}
\label{results-relevance-task}
Students tended to disclose sensitive information if they believed it was important for the chatbot to consider in generating its response. 
For example, for analyzing emails \textit{[A-T1]}, \emph{P10} shared \textit{names}, considering them essential to generate accurate summaries: \emph{``it's important for the names to be [in summaries] so I'd not anonymize it now.''} 
To find housing maintenance details \textit{[A-T2]}, \emph{P6} anonymized all items listed in the privacy panel, believing that none were relevant to the task: \emph{``Because none of those details [e.g., SSN, date of birth, physical address] should be relevant to any of the rules. 
That's like a matter of what the name is and email [...] that should be fine [to anonymize].''} 
When searching a car lease contract \textit{[B-T2]}, noticing \textit{dates of birth} in the panel prompted \emph{P7} to revisit the task description and the contract before deciding: \emph{``I think the dates of birth don't really matter [for this task], so I'll just retract.''}
For some participants, even a slight perceived impact of anonymization on chatbot accuracy was enough to forgo anonymization, as reflected in \emph{P9}'s comment on the chatbot's response for a task about finding utility information \textit{[B-T2]}: \emph{``So, the retraction doesn't seem to have affected the answer in any way, which is nice because I would not want [the chatbot] to lose accuracy in just protecting my own personal information.''}

\subsubsection{Sensitivity Level of Information (P1, P2, P7, P9).}
\label{results-sensitivity-level}
Students’ disclosure decisions were also influenced by how sensitive they perceived different information types to be.
When sharing a housing contract \textit{[A-T2]}, \emph{P1} chose not to remove a \textit{physical address} from their prompt, explaining: \emph{``[...] I don't consider [physical address] as a huge privacy concerning thing [...].''} 
In the same task, \emph{P2} immediately anonymized \textit{SSNs} using the panel before addressing any other flagged items. 
Similarly, while sharing a car lease agreement \textit{[B-T2]}, \emph{P9} noticed the \textit{SSN} flagged in the panel and reacted promptly: \emph{``SSN I definitely want to retract. I do not want my SSN going anywhere.''}
Students sometimes expressed uncertainty about what counts as sensitive. 
For example, while classifying customer complaints \textit{[B-T1]}, \emph{P7} questioned whether \textit{physical addresses} and \textit{dates of birth} should be treated as sensitive: \emph{``I didn't realize the address would be or date of birth would be that important. But I guess it is.''}

\subsubsection{Sensitive Information Ownership (P2, P4, P5, P7).}
Students sometimes treated information differently depending on whom it belongs to. 
When drafting an email \textit{[A-T3]}, \emph{P2} avoided sharing their own information (e.g., \textit{date of birth}, \textit{phone number}), explaining they are more cautious with their own personal details than with others': \emph{``[...] because this is my own contact detail, I'd be more skeptical about putting this as compared to someone else's, which is not good.''} 
In contrast, \emph{P4} shared their own information in the same task, noting: \emph{``This is also just my details, so I don't care [sharing].''}
When working with customer complaints data \textit{[B-T1]}, \emph{P5} considered the company's privacy preferences upon seeing the privacy panel, stating: \emph{``Task is identifying departments. Okay, so, this could be sensitive.''} 
Similarly, \emph{P7} expressed uncertainty about what would be considered sensitive for a financial advisor: \emph{``[...] I'm not a financial advisor actually, so, I'm not really sure what information would be the most sensitive.''}

\subsubsection{Convenience (P1, P5, P9).}
Students' disclosure decisions were also shaped by the time and effort required to protect information.
They weighed how easy it would be to identify and anonymize sensitive details, as \emph{P1} said: \emph{``I think if it's easy enough for me to cut [sensitive information] out when I search against it, I think I do that.''} 
In complaint classification task \textit{[B-T1]}, \emph{P5} explained that the panel made anonymization simpler than in their real-world experiences: \emph{``Normally, I wouldn't [anonymize], but because it has this feature, I think it's easier to just click [to anonymize] all of them [...].''}
\emph{P9} weighed the effort required to fix chatbot outputs when anonymization introduces inaccuracies: \emph{``I don't wanna give out fake names because then when I get the output back and it has fake names in it, I would've to manually again correct it back to the original names.''}
Greater convenience sometimes encouraged more cautious decisions when students were uncertain about information's sensitivity. For instance, \emph{P1} chose to anonymize \textit{names} flagged in the panel despite finding them acceptable to share, noting it was easy to do so: \emph{``I'm usually fine with [sharing names] but now that I have an option to do [anonymization] easier, I wouldn't mind retracting it.''}

\subsubsection{Partial Sharing to Reduce Risks (P2, P5).}
\label{results-partial-sharing}
Students viewed it as less risky to share only partial sensitive information about individuals. 
\emph{P2} explained this when searching a housing contract \textit{[A-T2]}: \emph{``It is like the combination of things that make it like phone number or date of birth [...] more risky to share. 
So, I think if I have at least just one [type of information], I think it is not enough to give much information about the person.''} 
Similarly, \emph{P5} copied an insurance document \textit{[B-T3]} starting just after their \textit{name} appeared at the top, noting: \emph{``As long as we just exclude the name, it should be fine.''}

\subsubsection{Public Availability of Sensitive Information (P1, P2).}
\label{results-public-availability}
Participants were more likely to share information they believed was already publicly available. 
\emph{P1}, for example, identified such information in a travel itinerary \textit{[A-T3]} and noted they do not mind sharing it: \emph{``I wouldn't consider any of this [...] very privacy concerning because all of this is like information that's available out there. All of these are names of venues and the places that will be visiting in the trip, which I don't mind sharing even if it were my own [...].''} 
\emph{P2} shared a similar view when noticing specific details in the chatbot's response: \emph{``I think this is not really all personal information because it's readily available, contact information and stuff. So, I think it's alright if I put it on a chat[bot].''}

\subsubsection{Use of Chatbot's Built-in Privacy Controls (P4).}
After \emph{P4} opted out of sharing content for model training and disabled the chatbot’s memory, they stopped using the privacy panel for anonymization, explaining: \emph{``[Chatbot owners] said they're not gonna train on [content], and they're gonna lose a lot of money if they actually train on it.''}

\subsection{Behaviors in Using Privacy-Protective Approaches Across Task-Completion Sessions (RQ3)}
\label{sec:rq3}
\subsubsection{Without Panel: No Use of Anonymization Strategies or Built-in Privacy Controls}
\label{sec:rq3_without}
When completing tasks using the chatbot \textit{without} the panel, \textit{none} of the participants used any anonymization strategies. Likewise, \textit{no one} looked for or used any of the built-in privacy controls (opting in/out of sharing content for model training, enabling/disabling memory), even though these controls were accessible from the profile icon, consistent with \textit{ChatGPT}'s interface design.

\subsubsection{With Panel: Varied Use of Anonymization Strategies Across and Within Information Types}
\label{results_varied_strategies}
In their use of the privacy panel, each participant varied their anonymization choices across types of sensitive information (except \emph{P1}, who exclusively used \textit{retracting}).
Likewise, all participants except \emph{P3} and \emph{P1} sometimes switched between approaches for a given information type (e.g., \textit{retracting} \textit{names} in one appearance of the panel and \textit{faking} them in another), indicating that their choices were not determined solely by information type.
Even so, every participant had at least one information type that they consistently anonymized using a single approach (e.g., always \textit{retracting} SSNs).
At a finer level, in each panel appearance, participants always applied the same anonymization approach to all instances of a given information type (e.g., \textit{generalizing} all instances of \textit{physical addresses}). This behavior occurred despite the panel allowing to apply a different approach to each instance, indicating that participants did not make instance-level adjustments.

\subsubsection{With Panel: Increased Awareness and Use of Built-in Privacy Controls}
\label{results_builtin_controls}
When using the chatbot \textit{with} the panel (which surfaced two \textit{ChatGPT}'s built-in privacy controls) eight participants (all except \textit{P8} and \textit{P10}) became aware of and explored these options through the panel, as evidenced by using the features, opening the related settings, or verbalizing their thoughts about them. 
Of these eight participants, five used at least one of the two built-in privacy controls.

\subsection{Rationale in Using or Avoiding Specific Privacy-Protective Approaches (RQ4)}
\label{sec:rq4}

When deciding which approaches to use, participants shared factors influencing their choices. We summarize these factors across anonymization strategies, built-in privacy controls, and an ad-hoc approach. We provide example quotes with task identifiers to clarify reasoning (e.g., \textit{[B-T1]} denotes the first task in \textit{Assignment B}, described in Appendix \ref{appendix_user_tasks}).

\subsubsection{Anonymization Approach > Retracting}
Students used \textit{retracting} as the highest level of anonymization when the identified information was highly sensitive (\emph{P2}, \emph{P9}) or task-irrelevant (\emph{P1}, \emph{P5}, \emph{P7}, \emph{P9}). 
For example, when looking for utility details in a housing contract \textit{[B-T2]}, \emph{P7} retracted all sensitive details listed in the panel (e.g., \textit{names}, \textit{SSNs}) because: \emph{``[...] for this purpose it's not really that important. I don't really need to know any of this stuff.''} 
Similarly, \emph{P1} retracted \textit{physical addresses} when searching a car contract \textit{[B-T2]}, saying: \emph{``I think [physical address] is irrelevant for answering the question.''} 
One reason for avoiding the \textit{retracting} approach was to allow the chatbot distinguish between instances of the same information type (\emph{P9}). \emph{P9} explained why they did not retract \textit{names} in the car contract: \emph{``[...][because chatbot] would need to be able to differentiate between my name and someone else's name. So, I guess I can keep the names.''}

\subsubsection{Anonymization Approach > Faking}
Students used the \textit{faking} approach when they wanted to provide more context than \textit{retracting} would allow (\emph{P4}, \emph{P5}). 
As \emph{P4} explained during the email classification task \textit{[A-T1]}: \emph{``I think I'd fake [email addresses] because this is email between people, so if you just retract, it might not have enough context.''} 
However, when students realized that faking replaces all instances of a type with the same dummy value, they avoided it to preserve distinctions between instances (\emph{P4}, \emph{P7}, \emph{P9}). 
\emph{P9} also worried that a fake value might unintentionally match someone else's real information: \emph{``Also, the problem with fake names is that they might actually be someone else's name.''}

\subsubsection{Anonymization Approach > Generalizing}
\textit{Generalizing} was available only for \textit{date of birth} and \textit{physical address}. 
Students chose this option when they believed the remaining details would still help the chatbot with the task (\emph{P2}, \emph{P4}, \emph{P6}, \emph{P9}) while being anonymized enough to prevent identification (\emph{P9}). For example, \emph{P6} generalized \textit{physical addresses} when looking for timing conflicts \textit{[A-T3]}, reasoning: \emph{``because it might need to know [physical addresses] to calculate the [...] travel distance between each of the different places and stuff.''} One reason for avoiding \textit{generalizing} was when multiple \textit{physical addresses} would have ended up looking identical (e.g., same state and zip code), as noted by P9: \emph{``a lot of them have different addresses but are of the same state. So, maybe generalization is not that good option here.''}

\subsubsection{Built-in Control > Opting Out of Sharing Content for Model Training}
\label{rq2-opt-out}
Students shared differing views and made varied choices when noticing this option. Some opted out (\emph{P1}, \emph{P4}, \emph{P6}, \emph{P7}), viewing it as an easy way to protect their information from future uses (\emph{P1}, \emph{P6}, \emph{P7}), as \emph{P1} explained: \emph{``I would rather opt out. This is an easier way to ensure that my information is not being used in any way.''} However, level of confidence in this feature varied. While \emph{P4} trusted it: \emph{``[...] they promise not to train on your data. I trust them for that''}, \emph{P1} doubted its effectiveness:\emph{``I'm not fully confident that just, opting out will ensure that anything private that I might share would not be used in any way without my consent.''}
Other students kept the default setting---that is, opting in (\emph{P2}, \emph{P3}, \emph{P5}, \emph{P9}, \emph{P10}). Most did not show any reasoning, but \emph{P9} noted they had already anonymized most sensitive details using the privacy panel: \emph{``I don't mind sharing content for model training cause I've retracted most of the important information.''} When first encountering this option in the panel, \emph{P3} commented: \emph{``I would opt out of this''}, but after reading its description in the pop-up window (consistent with ChatGPT's description, Appendix \ref{appendix_interface_without} Figure \ref{fig:settings-panel}), they decided to continue sharing their content for model training.

\subsubsection{Built-in Control > Disabling Memory}
\label{rq2-memory}
Participants were more inclined to keep memory enabled, citing two benefits: easier prompting (e.g., reusing a single prompt across interactions) and improved response quality (\emph{P1}, \emph{P2}, \emph{P3}, \emph{P5}, \emph{P6}, \emph{P7}, \emph{P9}).
For example, \emph{P2} highlighted the convenience of reusing prompts: \emph{``I think it's easier to put the prompt once and continue that chat with different tasks for the same prompt which wouldn't be possible if I disable the chatbot's memory''}, and \emph{P5} emphasized memory's potential positive effect on the chatbot's performance: \emph{``I felt as if [memory] would significantly impact the chatbot's function and ability to complete the task [...]''}. 
Only two participants disabled memory in their second session (when using the chatbot \textit{with} the panel): \emph{P2} in their third task and \emph{P4} in their first task. Both showing uncertainty about how disabling memory would function. \emph{P2} noted their doubts directly: \emph{``I can [disable memory] at least with this task. But I wonder if it means it won't remember the previous messages also''}, while \emph{P4}’s subsequent interactions showed they had not expected the chatbot to forget past messages within the same conversation.

\subsubsection{Prompting Strategy}
An ad-hoc approach we observed was the use of \textit{prompting strategies} to ask the chatbot to protect sensitive data. When \emph{P4} encountered issues with the faking approach---all instances being replaced with the same value---during the email analysis task \textit{[A-T1]}, they instead wrote explicit prompts instructing the chatbot to replace names, such as: \emph{``Replace everyone's name to protect identity [...]''} and \emph{``Replace the name in the emails to make it secure.''}

\section{Discussion and Future Work}
\label{sec:discussion}
Our findings suggest that user-facing privacy tools can increase awareness of sensitive details in prompts and promote reasoning about and engagement in protective actions. 
We synthesize our findings across RQs and discuss opportunities to further enhance privacy engagement and support by targeting key decision-making phases.

\subsection{‌Bringing Privacy into Users' Attentional Focus}
When participants used the chatbot to complete tasks, they needed to navigate uncertainties in refining prompts, provide sufficient context, and evaluate the accuracy of chatbot responses. 
Within and across these tasks, interactions that threaten privacy can be easily overlooked. 
Indeed, when our participants used the chatbot \textit{without} the panel, \textit{8 out of 10} showed no privacy considerations, and \textit{none} used any protective approaches (Sections \ref{sec:rq1_without} and \ref{sec:rq3_without}).
In contrast, our panel raised privacy awareness at critical moments of disclosure and encouraged users to pause and weigh the costs and benefits of sharing, prompting them to take protective actions.
This impact was evident in participants’ reflections; as \emph{P3} explained: \emph{``This study made me more aware of [sensitive information] and made me more mindful about how I interact with [the chatbot].''}
and \emph{P5} noted: \emph{``I usually take prompts and information and just paste them in order to ask chatbots to do things. However, I usually don't use privacy features, but the chatbot provided [in this study,] had this feature easily accessible, so I decided to use it.''}
Our panel was designed for optional use, but future work can examine how varying levels of enforcement \cite{willermark-2020,Buono-2023} (e.g., requiring panel interaction before proceeding) affect privacy awareness and reasoning. 
Future research might also explore how users' existing cognitive efforts shift when privacy tools intercept chatbot interactions \cite{lev-2024}---e.g., whether increased attention to privacy reduces efforts for evaluating responses.

For built-in privacy controls, surfacing them by the panel and framing them as privacy-protective actions enhanced their discoverability and use (Section \ref{results_builtin_controls}).
Yet, participants remained uncertain about the effectiveness of these controls in protecting privacy (Sections \ref{rq2-opt-out} and \ref{rq2-memory}).
This emphasizes a need for making these features more visible---rather than relying on users' discovery---and transparently communicating how they function, the protections they provide, and the trade-offs involved \cite{zhang-2024,malki2025hoovered}. 
For \textit{ChatGPT}, however, the framing of the opt-out feature emphasizes the benefits of opting in, omitting any immediate mentions of privacy risks (Appendix \ref{appendix_interface_without}, Figure \ref{fig:settings-panel}). 
Similarly, the memory option can be explained more transparently, particularly for its impact on privacy and response quality (Section \ref{rq2-memory}).

\subsection{Supporting Users in Their Privacy Reasoning}
Our analysis showed that encountering the panel encouraged engagement in contextual reasoning about disclosure and protection.
However, \textbf{students' rationale revealed potential flawed or outdated conceptions}. 
Prior work highlights that users may not know what types of information are sufficient to identify individuals \cite{song-2025}, may underestimate the extent to which LLM-based CAs can connect data across sources \cite{lee-2024}, and may hold oversimplified or inaccurate mental models of how such systems work \cite{zhang-2024}.
Some of our participants showed uncertainty about what counts as sensitive and assumed that sharing partial or publicly available information carries limited risk (Sections \ref{results-sensitivity-level}, \ref{results-partial-sharing}, \ref{results-public-availability}). 
Yet they may not realize that modern AI systems can automate large-scale, real-time cross-referencing and inference, extending far beyond the content explicitly disclosed in a conversation \cite{staab-2023,lee-2024}. 
This gap in understanding is reflected in \emph{P7}’s comment: \emph{``I knew sharing sensitive information wasn't probably the best idea, but I wasn't aware of the possibility of the information being linked to a larger digital presence/contributing to a larger pool of collected data already.''} 
Such disconnects can amplify the disclosure of sensitive information.
Thus, we encourage future studies to explore ways of addressing gaps in users’ privacy knowledge and literacy both through their interactions with CAs \cite{Zhou-2025-rescriber,Chen-2025-CLEAR} and through standalone educational tools. 
Standalone tools can help with further examining and deepening users’ understanding of how CAs operate and process data, the risks of data leakage or misuse across processing stages, and the possibility of real-world AI failures by illustrating past privacy incidents involving CAs \cite{feffer-2023,hadinezhad-2025,castro-2025-using}. 
These approaches can foster more critical reflection and help users better balance immediate benefits with the potential long-term disclosure risks.

\subsection{Supporting Users in Filtering Task-Irrelevant Sensitive Information}
We found that students often disclosed sensitive information and avoided certain protective approaches to provide more context and improve chatbot response quality (Sections \ref{results-relevance-task} and \ref{sec:rq4}), aligning with prior work's findings \cite{zhang-2024,Zhou-2025-rescriber,Brown-2022-whatdoes}. 
However, \textbf{during interactions, it was difficult for users to assess how sharing certain information and the level of specificity of the information to be shared (retracted, generalized, faked), affects response quality}.
This difficulty can lead to prioritizing chatbot utility over privacy, a pattern we repeatedly observed.
We therefore encourage developing additional supports in filtering task-irrelevant details. 
One approach is for CAs to scaffold incremental prompting by explicitly requesting only necessary inputs rather than keeping the requests open-ended. 
For example, when \emph{P6} sent a message saying they would share emails for sentiment classification, the chatbot could have requested only the message bodies and prompted withholding sender details (Appendix \ref{appendix_discussion}, Figure \ref{fig:incremental_prompting}). 
Future systems could also leverage locally deployed lightweight language models (e.g., Edge LLMs \cite{zheng-2025-a-review,corradini-2025}) to analyze task context in real time and suggest removing irrelevant and/or sensitive content.

\subsection{Balancing Automation and User Control in Protecting Sensitive Information}
Our analysis indicates opportunities for automating efforts for protecting sensitive information.
User control remains essential where decisions depend on contextual factors (e.g., data ownership, broader task goals) or personal preferences (e.g., how sensitive certain details are to an individual) that users possess but that may not, or should not, be available to CAs.
Automation, however, can still support users in other areas, such as detecting sensitive information in messages \cite{Zhou-2025-rescriber, Chen-2025-CLEAR,Chong-2025-casper} or replacing anonymized values with original values in chatbot responses (e.g., via a temporary mapping between original and masked values) \cite{Zhou-2025-rescriber}.
In addition, some contextual factors may change slowly enough to enable some personalized automation.
For instance, we found that each participant consistently anonymized at least one information type using the same strategy (e.g., always \textit{retracting} SSNs) (Section \ref{results_varied_strategies}). 
This suggests an opportunity in allowing users to specify preferences---e.g., automatically applying a chosen strategy to certain information types, being prompted before submitting others, or never anonymizing some.
Two considerations arise: first, such preferences must be stored securely and allow for complete deletion. 
Second, user preferences may be unstable or only partially informed; 
they can shift as users encounter new contexts (see theories of temporal decision-making \cite{harris-2022} and preference reversal \cite{johnson-1988}). 
This was evident in the interactions and comments of \emph{P9} and \emph{P5}.
For example, although \emph{P9} initially suggested a universal “anonymize all” feature: \emph{``There should maybe be one [button] at the top that says [...] retract all of these or generalize all of these or fake all of these. So I can just select that and like have it apply to all [...], but just with the click of one button''}, their subsequent behavior showed mixed and situational use of anonymization strategies. 
\textbf{Thus, automation should be guided by user-defined preferences while also continuously making users aware of contextual factors that may shape their preferences over time.}

\subsection{Promoting Manual Protections Where Privacy Tools Are Limited}
User-facing privacy tools may not be comprehensive in all aspects. 
For instance, tools that identify sensitive information in user messages may fail to detect all such details, making transparent communication of these limitations essential.
Our panel included a brief note informing participants of this constraint (Figure \ref{fig:with_panel}, bottom of B).
While \emph{P2} and \emph{P7} appreciated this transparency, [\emph{P2}]: \emph{``I liked that in the panel at the end, the note also said that all these [information] might not be all the sensitive info in the prompt so the user can scan through the prompt themselves too''}, the note alone was insufficient to encourage manual efforts, as we observed very limited protection of unflagged sensitive information (Section \ref{results_unflagged_info_behavior}). 
\textbf{This suggests a need for future work to develop more effective just-in-time communication of tool limitations that encourages user actions in filling tool's gaps} \cite{henestrosa-2025,nngroup_explainable,hadi-2024-math}. 
Researchers can examine \textit{when} such communication is most effective (e.g., when messages likely contain unflagged sensitive details), \textit{where} it should appear (e.g., as standalone warnings or embedded within existing privacy tools), and \textit{what} it should convey (e.g., actionable guidance, examples of commonly missed information, or explanations of tools' detection capabilities).

\section{Limitations}
While our simulation captured key elements of \textit{ChatGPT}’s interface, it was not an exact replica. For example, API responses were displayed with basic visual formatting, which (despite being explained to participants) may have affected interactions (e.g., making math-related responses harder to interpret).
Future work can also extend sensitive information detection to additional content (e.g., health or financial details). 
We implemented a simple memory feature that stores only recent messages, while the setting panel mirrored \textit{ChatGPT’s} descriptions (Appendix \ref{appendix_interface_without}, Figure \ref{fig:settings-panel}). 
Although we briefed participants on how the memory works, it may still have caused confusion. 
Lastly, we focused on \textit{ChatGPT} due to its widespread use; however, future studies can explore similar privacy panels in other CAs and include broader populations across communities and geographic contexts to capture broader privacy behaviors and design needs.

\section{Conclusion}
We discussed students' disclosure and protection behaviors when using a simulated ChatGPT interface with and without a just-in-time privacy notice panel as well as the reasoning underlying these behaviors.
Our findings---drawn from user data across task completions and surveys---suggested that the panel effectively raised privacy awareness at critical moments of disclosure, encouraged participants to pause and reason about what to share or withhold, and supported them in applying protective actions.
In particular, students drew on a variety of contextual factors when making these decisions, and they were less likely to disclose sensitive information when using the chatbot integrated with the panel.
These results suggest that interface-level tools can meaningfully empower users to take proactive protective measures and make certain privacy risks impractical from the outset, while still benefiting from capabilities of CAs. 
These may also make it easier for CA owners and developers to safeguard sensitive data.  

\begin{acks}

\end{acks}

\bibliographystyle{ACM-Reference-Format}
\bibliography{02_references}

@inproceedings{zhang-2024,
  author    = {Zhang, Zhiping and Jia, Michelle and Lee, Hao-Ping (Hank) and Yao, Bingsheng and Das, Sauvik and Lerner, Ada and Wang, Dakuo and Li, Tianshi},
  title     = {``It's a fair game,'' or is it? Examining how users navigate disclosure risks and benefits when using {LLM}-based conversational agents},
  booktitle = {Proceedings of the 2024 CHI Conference on Human Factors in Computing Systems},
  year      = {2024},
  doi       = {10.1145/3613904.3642385},
  numpages  = {26}
}

@inproceedings{song-2025,
  author    = {Song, Qiurong and Wu, Yanlai and Hernandez, Rie Helene (Lindy) and Li, Yao and Kou, Yubo and Gui, Xinning},
  title     = {Understanding users' perception of personally identifiable information},
  booktitle = {Proceedings of the 2025 CHI Conference on Human Factors in Computing Systems},
  year      = {2025},
  doi       = {10.1145/3706598.3713783},
  numpages  = {24}
}

@inproceedings{nasr-2025,
  author    = {Nasr, Milad and Rando, Javier and Carlini, Nicholas and Hayase, Jonathan and Jagielski, Matthew and Cooper, A. Feder and Ippolito, Daphne and Choquette-Choo, Christopher A. and Tram{\`e}r, Florian and Lee, Katherine},
  title     = {Scalable extraction of training data from aligned, production language models},
  booktitle = {Proceedings of the Thirteenth International Conference on Learning Representations},
  year      = {2025},
  url       = {https://openreview.net/forum?id=vjel3nWP2a}
}

@inproceedings{carlini-2023,
  author    = {Carlini, Nicholas and Ippolito, Daphne and Jagielski, Matthew and Lee, Katherine and Tram{\`e}r, Florian and Zhang, Chiyuan},
  title     = {Quantifying memorization across neural language models},
  booktitle = {Proceedings of the Eleventh International Conference on Learning Representations},
  year      = {2023},
  url       = {https://openreview.net/forum?id=TatRHT_1cK}
}

@inproceedings{lee-2024,
  author    = {Lee, Hao-Ping (Hank) and Yang, Yu-Ju and von Davier, Thomas Serban and Forlizzi, Jodi and Das, Sauvik},
  title     = {Deepfakes, phrenology, surveillance, and more: A taxonomy of {AI} privacy risks},
  booktitle = {Proceedings of the 2024 CHI Conference on Human Factors in Computing Systems},
  year      = {2024},
  doi       = {10.1145/3613904.3642116},
  numpages  = {19}
}

@misc{AIAAIC-LeeLuda,
  author       = {{AIAAIC}},
  title        = {Lee Luda AI chatbot spouts offensive responses},
  year         = {2021},
  howpublished = {Online; AIAAIC Repository: AI, Algorithmic and Automation Incidents and Controversies},
  note         = {Accessed: 2026-01-13},
  url          = {https://www.aiaaic.org/aiaaic-repository/ai-algorithmic-and-automation-incidents/lee-luda-chatbot}
}

@misc{AIAAIC-bard,
  author       = {{AIAAIC}},
  title        = {Google Search indexes Bard personal chats},
  year         = {2023},
  howpublished = {Online; AIAAIC Repository: AI, Algorithmic and Automation Incidents and Controversies},
  note         = {Accessed: 2026-01-13},
  url          = {https://www.aiaaic.org/aiaaic-repository/ai-algorithmic-and-automation-incidents/google-search-indexes-bard-personal-chats}
}

@inproceedings{yu-2024,
  author    = {Yu, Da and Kairouz, Peter and Oh, Sewoong and Xu, Zheng},
  title     = {Privacy-preserving instructions for aligning large language models},
  booktitle = {Proceedings of the 41st International Conference on Machine Learning},
  year      = {2024},
  numpages  = {27}
}

@article{kandpal-2022,
  title={Deduplicating Training Data Mitigates Privacy Risks in Language Models},
  author={Nikhil Kandpal and Eric Wallace and Colin Raffel},
  journal={ArXiv},
  year={2022},
  url={https://api.semanticscholar.org/CorpusID:246823128}
}

@inproceedings{jang-2023-knowledge,
  author    = {Jang, Joel and Yoon, Dongkeun and Yang, Sohee and Cha, Sungmin and Lee, Moontae and Logeswaran, Lajanugen and Seo, Minjoon},
  title     = {Knowledge unlearning for mitigating privacy risks in language models},
  booktitle = {Proceedings of the 61st Annual Meeting of the Association for Computational Linguistics (Volume 1: Long Papers)},
  year      = {2023},
  pages     = {14389--14408},
  doi       = {10.18653/v1/2023.acl-long.805}
}

@article{ait-mlouk-2023,
  author  = {Ait-Mlouk, Addi and Alawadi, Sadi and Toor, Salman and Hellander, Andreas},
  title   = {FedBot: Enhancing privacy in chatbots with federated learning},
  journal = {arXiv preprint},
  year    = {2023},
  url     = {https://arxiv.org/abs/2304.03228}
}

@inproceedings{bagdasarian-2024,
  author    = {Bagdasarian, Eugene and Yi, Ren and Ghalebikesabi, Sahra and Kairouz, Peter and Gruteser, Marco and Oh, Sewoong and Balle, Borja and Ramage, Daniel},
  title     = {AirGapAgent: Protecting privacy-conscious conversational agents},
  booktitle = {Proceedings of the 2024 ACM SIGSAC Conference on Computer and Communications Security},
  year      = {2024},
  pages     = {3868--3882},
  doi       = {10.1145/3658644.3690350}
}

@article{majmudar-2022,
  author  = {Majmudar, Jimit and Dupuy, Christophe and Peris, Charith and Smaili, Sami and Gupta, Rahul and Zemel, Richard},
  title   = {Differentially private decoding in large language models},
  journal = {arXiv preprint},
  year    = {2022},
  url     = {https://www.amazon.science/publications/differentially-private-decoding-in-large-language-models}
}

@inproceedings{mutahar-2025,
  author    = {Ali, Mutahar and Arunasalam, Arjun and Farrukh, Habiba},
  title     = {Understanding users' security and privacy concerns and attitudes towards conversational {AI} platforms},
  booktitle = {Proceedings of the 2025 IEEE Symposium on Security and Privacy},
  year      = {2025},
  pages     = {298--316},
  doi       = {10.1109/SP61157.2025.00241}
}

@article{nissenbaum-2004,
  author  = {Nissenbaum, Helen},
  title   = {Privacy as contextual integrity},
  journal = {Washington Law Review},
  volume  = {79},
  year    = {2004},
  url     = {https://digitalcommons.law.uw.edu/wlr/vol79/iss1/10},

}

@article{joinson-2001,
  author  = {Joinson, Adam N.},
  title   = {Self-disclosure in computer-mediated communication: The role of self-awareness and visual anonymity},
  journal = {European Journal of Social Psychology},
  volume  = {31},
  number  = {2},
  pages   = {177--192},
  year    = {2001},
  doi     = {10.1002/ejsp.36}
}

@article{reis-2017,
  author  = {Reis, Harry T. and Lemay Jr., Edward P. and Finkenauer, Catrin},
  title   = {Toward understanding understanding: The importance of feeling understood in relationships},
  journal = {Social and Personality Psychology Compass},
  volume  = {11},
  number  = {3},
  year    = {2017},
  doi     = {10.1111/spc3.12308}
}

@article{croes-2024,
  author  = {Croes, Emmelyn A. J. and Antheunis, Marjolijn L. and van der Lee, Chris and de Wit, Jan M. S.},
  title   = {Digital confessions: The willingness to disclose intimate information to a chatbot and its impact on emotional well-being},
  journal = {Interacting with Computers},
  volume  = {36},
  number  = {5},
  pages   = {279--292},
  year    = {2024},
  doi     = {10.1093/iwc/iwae016}
}

@article{ho-annabell-2018,
  author  = {Ho, Annabell and Hancock, Jeff and Miner, Adam S.},
  title   = {Psychological, relational, and emotional effects of self-disclosure after conversations with a chatbot},
  journal = {Journal of Communication},
  volume  = {68},
  number  = {4},
  pages   = {712--733},
  year    = {2018},
  doi     = {10.1093/joc/jqy026}
}

@article{johnson-1988,
  author  = {Johnson, Eric J. and Payne, John W. and Bettman, James R.},
  title   = {Information displays and preference reversals},
  journal = {Organizational Behavior and Human Decision Processes},
  volume  = {42},
  number  = {1},
  pages   = {1--21},
  year    = {1988},
  doi     = {10.1016/0749-5978(88)90017-9}
}

@article{harris-2022,
  author  = {Harris, Alison and Hutcherson, Cendri A.},
  title   = {Temporal dynamics of decision making: A synthesis of computational and neurophysiological approaches},
  journal = {WIREs Cognitive Science},
  volume  = {13},
  number  = {3},
  year    = {2022},
  doi     = {10.1002/wcs.1586}
}

@inproceedings{feffer-2023,
  author    = {Feffer, Michael and Martelaro, Nikolas and Heidari, Hoda},
  title     = {The {AI} Incident Database as an educational tool to raise awareness of {AI} harms},
  booktitle = {Proceedings of the 3rd ACM Conference on Equity and Access in Algorithms, Mechanisms, and Optimization},
  year      = {2023},
  numpages  = {11},
  doi       = {10.1145/3617694.3623223}
}

@inproceedings{hadinezhad-2025,
  author    = {Hadi Nezhad, Mohammad and Castro, Francisco Enrique Vicente and Mak, Eugene and Haas, Peter J. and Allessio, Danielle and Osterweil, Leon and Rasul, Injila and Conboy, Heather and Arroyo, Ivon},
  title     = {Embedding ethical awareness in computer science and {AI} education: The {PEaRCE} approach to responsible computing},
  booktitle = {Proceedings of the 26th International Conference on Artificial Intelligence in Education},
  year      = {2025},
  pages     = {135--149},
  doi       = {10.1007/978-3-031-98414-3_10}
}

@article{zhang-2025-exploring,
  author  = {Zhang, Xing and Li, Zhaoqian and Zhang, Mingyang and Yin, Mingyue and Yang, Zhangyu and Gao, Dong and Li, Hansen},
  title   = {Exploring artificial intelligence ({AI}) chatbot usage behaviors and their association with mental health outcomes in Chinese university students},
  journal = {Journal of Affective Disorders},
  volume  = {380},
  pages   = {394--400},
  year    = {2025},
  doi     = {10.1016/j.jad.2025.03.141}
}

@article{stohr-2024,
  author  = {St{\"o}hr, Christian and Ou, Amy Wanyu and Malmstr{\"o}m, Hans},
  title   = {Perceptions and usage of {AI} chatbots among students in higher education across genders, academic levels and fields of study},
  journal = {Computers and Education: Artificial Intelligence},
  volume  = {7},
  pages   = {100259},
  year    = {2024},
  doi     = {10.1016/j.caeai.2024.100259}
}

@article{lappeman-2022-trust,
  author  = {Lappeman, James and Marlie, Siddeeqah and Johnson, Tamryn and Poggenpoel, Sloane},
  title   = {Trust and digital privacy: Willingness to disclose personal information to banking chatbot services},
  journal = {Journal of Financial Services Marketing},
  volume  = {28},
  number  = {2},
  year    = {2022},
  doi     = {10.1057/s41264-022-00154-z}
}

@inproceedings{esmaeili-2024,
  author    = {Esmaeili, Mona and Ahmadi, Mohammad and Ismaeil, Mohammad David and Mirzaei, Sharareh and Canales Verdial, Jorge},
  title     = {Advancements in {AI}-driven customer service},
  booktitle = {Proceedings of the 2024 IEEE World {AI} {IoT} Congress (AIIoT)},
  year      = {2024},
  pages     = {1--5},
  doi       = {10.1109/AIIoT61789.2024.10579008}
}

@inproceedings{kelley-2023,
  author    = {Kelley, Patrick Gage and Cornejo, Celestina and Hayes, Lisa and Jin, Ellie Shuo and Sedley, Aaron and Thomas, Kurt and Yang, Yongwei and Woodruff, Allison},
  title     = {``There will be less privacy, of course'': How and why people in 10 countries expect {AI} will affect privacy in the future},
  booktitle = {Proceedings of the Nineteenth Symposium on Usable Privacy and Security (SOUPS 2023)},
  year      = {2023},
  pages     = {579--603},
  url       ={https://www.usenix.org/conference/soups2023/presentation/kelley},
}

@article{braun-2006,
  author  = {Braun, Virginia and Clarke, Victoria},
  title   = {Using thematic analysis in psychology},
  journal = {Qualitative Research in Psychology},
  volume  = {3},
  number  = {2},
  pages   = {77--101},
  year    = {2006},
  doi     = {10.1191/1478088706qp063oa}
}

@article{mireshghallah-2024,
  author  = {Mireshghallah, Niloofar and Antoniak, Maria and More, Yash and Choi, Yejin and Farnadi, Golnoosh},
  title   = {Trust no bot: Discovering personal disclosures in human-{LLM} conversations in the wild},
  journal = {arXiv preprint},
  year    = {2024},
  url     = {https://arxiv.org/abs/2407.11438}
}

@article{corradini-2025,
  author  = {Corradini, Flavio and Leonesi, Matteo and Piangerelli, Marco},
  title   = {State of the art and future directions of small language models: A systematic review},
  journal = {Big Data and Cognitive Computing},
  volume  = {9},
  number  = {7},
  year    = {2025},
  doi     = {10.3390/bdcc9070189}
}

@article{zheng-2025-a-review,
  author  = {Zheng, Yue and Chen, Yuhao and Qian, Bin and Shi, Xiufang and Shu, Yuanchao and Chen, Jiming},
  title   = {A review on edge large language models: Design, execution, and applications},
  journal = {ACM Computing Surveys},
  volume  = {57},
  number  = {8},
  year    = {2025},
  doi     = {10.1145/3719664},
  numpages = {35}
}

@inproceedings{Zhou-2025-rescriber,
  author    = {Zhou, Jijie and Xu, Eryue and Wu, Yaoyao and Li, Tianshi},
  title     = {Rescriber: Smaller-{LLM}-powered user-led data minimization for {LLM}-based chatbots},
  booktitle = {Proceedings of the 2025 CHI Conference on Human Factors in Computing Systems},
  year      = {2025},
  doi       = {10.1145/3706598.3713701},
  numpages  = {28}
}

@inproceedings{Chen-2025-CLEAR,
  author    = {Chen, Chaoran and Zhou, Daodao and Ye, Yanfang and Li, Toby Jia-Jun and Yao, Yaxing},
  title     = {{CLEAR}: Towards contextual {LLM}-empowered privacy policy analysis and risk generation for large language model applications},
  booktitle = {Proceedings of the 30th International Conference on Intelligent User Interfaces},
  year      = {2025},
  pages     = {277--297},
  doi       = {10.1145/3708359.3712156}
}

@inproceedings{Chong-2025-casper,
  author    = {Chong, Chun Jie and Hou, Chenxi and Yao, Zhihao and Talebi, Seyed Mohammadjavad Seyed},
  title     = {Casper: Prompt sanitization for protecting user privacy in web-based large language models},
  booktitle = {Proceedings of the 2025 IEEE 12th International Conference on Cyber Security and Cloud Computing (CSCloud)},
  year      = {2025},
  pages     = {122--133},
  doi       = {10.1109/CSCloud66326.2025.00027}
}

@article{schaar-2010,
  author  = {Schaar, Peter},
  title   = {Privacy by design},
  journal = {Identity in the Information Society},
  volume  = {3},
  number  = {2},
  pages   = {267--274},
  year    = {2010},
  doi     = {10.1007/s12394-010-0055-x}
}

@inproceedings{lev-2024,
  author    = {Tankelevitch, Lev and Kewenig, Viktor and Simkute, Auste and Scott, Ava Elizabeth and Sarkar, Advait and Sellen, Abigail and Rintel, Sean},
  title     = {The metacognitive demands and opportunities of generative {AI}},
  booktitle = {Proceedings of the 2024 CHI Conference on Human Factors in Computing Systems},
  year      = {2024},
  doi       = {10.1145/3613904.3642902},
  numpages  = {24}
}

@article{staab-2023,
  author  = {Staab, Robin and Vero, Mark and Balunovi{\'c}, Mislav and Vechev, Martin},
  title   = {Beyond memorization: Violating privacy via inference with large language models},
  journal = {arXiv preprint},
  year    = {2023},
  url     = {https://arxiv.org/abs/2310.07298}
}

@inproceedings{castro-2025-using,
  author    = {Castro, Francisco Enrique Vicente and Hadi Nezhad, Mohammad and Iyer, Mark and Arroyo, Ivon},
  title     = {Using open-ended responses to design an assessment tool for ethical and responsible computing learning activities},
  booktitle = {Proceedings of the 25th Koli Calling International Conference on Computing Education Research},
  year      = {2025},
  doi       = {10.1145/3769994.3770060},
  numpages  = {3}
}

@inproceedings{hadi-2024-math,
  author    = {Hadi Nezhad, Mohammad and Castro, Francisco and Woolf, Beverly and Arroyo, Ivon},
  title     = {Math teachers' in-class information needs and usage for effective design of classroom orchestration tools},
  booktitle = {Proceedings of the European Conference on Technology Enhanced Learning},
  year      = {2024},
  pages     = {299--314},
  doi       = {10.1007/978-3-031-72315-5_21}
}

@inproceedings{malki2025hoovered,
  author    = {Malki, Lisa Mekioussa and others},
  title     = {``Hoovered up as a data point'': Exploring privacy behaviours, awareness, and concerns among {UK} users of {LLM}-based conversational agents},
  booktitle = {Proceedings on Privacy Enhancing Technologies},
  year      = {2025},
  doi       = {10.56553/popets-2025-0160}
}

@inproceedings{Brown-2022-whatdoes,
  author    = {Brown, Hannah and Lee, Katherine and Mireshghallah, Fatemehsadat and Shokri, Reza and Tram{\`e}r, Florian},
  title     = {What does it mean for a language model to preserve privacy?},
  booktitle = {Proceedings of the 2022 ACM Conference on Fairness, Accountability, and Transparency},
  year      = {2022},
  pages     = {2280--2292},
  doi       = {10.1145/3531146.3534642}
}

@article{zhang2024adanonymizer,
  author  = {Zhang, Shuning and Yi, Xin and Xing, Haobin and Ye, Lyumanshan and Hu, Yongquan and Li, Hewu},
  title   = {Adanonymizer: Interactively navigating and balancing the duality of privacy and output performance in human-{LLM} interaction},
  journal = {arXiv preprint},
  year    = {2024},
  url     = {https://arxiv.org/abs/2410.15044}
}

@inproceedings{2024-maeda,
  author    = {Maeda, Takuya and Quan-Haase, Anabel},
  title     = {When human-{AI} interactions become parasocial: Agency and anthropomorphism in affective design},
  booktitle = {Proceedings of the 2024 ACM Conference on Fairness, Accountability, and Transparency},
  year      = {2024},
  pages     = {1068--1077},
  doi       = {10.1145/3630106.3658956}
}

@inproceedings{2020-Ischen,
  author    = {Ischen, Carolin and Araujo, Theo and Voorveld, Hilde and van Noort, Guda and Smit, Edith},
  title     = {Privacy concerns in chatbot interactions},
  booktitle = {Chatbot Research and Design},
  year      = {2020},
  pages     = {34--48},
  doi       = {10.1007/978-3-030-39540-7_3}
}

@article{gabriel2024ethics,
  title={The ethics of advanced ai assistants},
  author={Gabriel, Iason and Manzini, Arianna and Keeling, Geoff and Hendricks, Lisa Anne and Rieser, Verena and Iqbal, Hasan and Toma{\v{s}}ev, Nenad and Ktena, Ira and Kenton, Zachary and Rodriguez, Mikel and others},
  journal={arXiv preprint arXiv:2404.16244},
  year={2024}
}

@inproceedings{mutahar-2025-understanding,
  author    = {Ali, Mutahar and Arunasalam, Arjun and Farrukh, Habiba},
  title     = {Understanding users' security and privacy concerns and attitudes towards conversational {AI} platforms},
  booktitle = {Proceedings of the 2025 IEEE Symposium on Security and Privacy},
  year      = {2025},
  pages     = {298--316},
  doi       = {10.1109/SP61157.2025.00241}
}

@inproceedings{naeini2017privacy,
  author    = {Naeini, Pardis Emami and Bhagavatula, Sruti and Habib, Hana and Degeling, Martin and Bauer, Lujo and Cranor, Lorrie Faith and Sadeh, Norman},
  title     = {Privacy expectations and preferences in an {IoT} world},
  booktitle = {Proceedings of the Thirteenth Symposium on Usable Privacy and Security (SOUPS 2017)},
  year      = {2017},
  pages     = {399--412},
url = {https://www.usenix.org/conference/soups2017/technical-sessions/presentation/naeini}
}

@techreport{NBER-2025,
  author      = {Chatterji, Aaron and Cunningham, Thomas and Deming, David J. and Hitzig, Zoe and Ong, Christopher and Shan, Carl Yan and Wadman, Kevin},
  title       = {How people use {ChatGPT}},
  institution = {National Bureau of Economic Research},
  type        = {Working Paper},
  number      = {34255},
  year        = {2025},
  doi         = {10.3386/w34255}
}

@article{henestrosa-2025,
  author  = {Henestrosa, Angelica Lermann and Kimmerle, Joachim},
  title   = {``Always check important information!'': The role of disclaimers in the perception of {AI}-generated content},
  journal = {Computers in Human Behavior: Artificial Humans},
  volume  = {4},
  pages   = {100142},
  year    = {2025},
  doi     = {10.1016/j.chbah.2025.100142}
}

@misc{nngroup_explainable,
  author       = {{Nielsen Norman Group}},
  title        = {Explainable AI: Definition, Benefits, and Examples},
  year         = {2023},
  howpublished = {\url{https://www.nngroup.com/articles/explainable-ai/}},
  note         = {Accessed: 2026-01-14}
}

@inproceedings{willermark-2020,
  author    = {Willermark, Sara and {\'I}slind, Anna Sigr{\'\i}{\dh}ur},
  title     = {The polite pop-up: An experimental study of pop-up design characteristics and user experience},
  booktitle = {Proceedings of the 53rd Hawaii International Conference on System Sciences},
  year      = {2020},
  doi       = {10.24251/HICSS.2020.514}
}

@inproceedings{Buono-2023,
  author    = {Buono, Paolo and Desolda, Giuseppe and Greco, Francesco and Piccinno, Antonio},
  title     = {Let warnings interrupt the interaction and explain: Designing and evaluating phishing email warnings},
  booktitle = {Extended Abstracts of the CHI Conference on Human Factors in Computing Systems},
  year      = {2023},
  doi       = {10.1145/3544549.3585802},
  numpages  = {6}
}

\appendix

\section{Supplementary Methodological Materials}
\subsection{ChatGPT Interface Simulation}
\label{appendix_interface_without}
\begin{figure}[h]
  \centering
  \includegraphics[width=\linewidth]{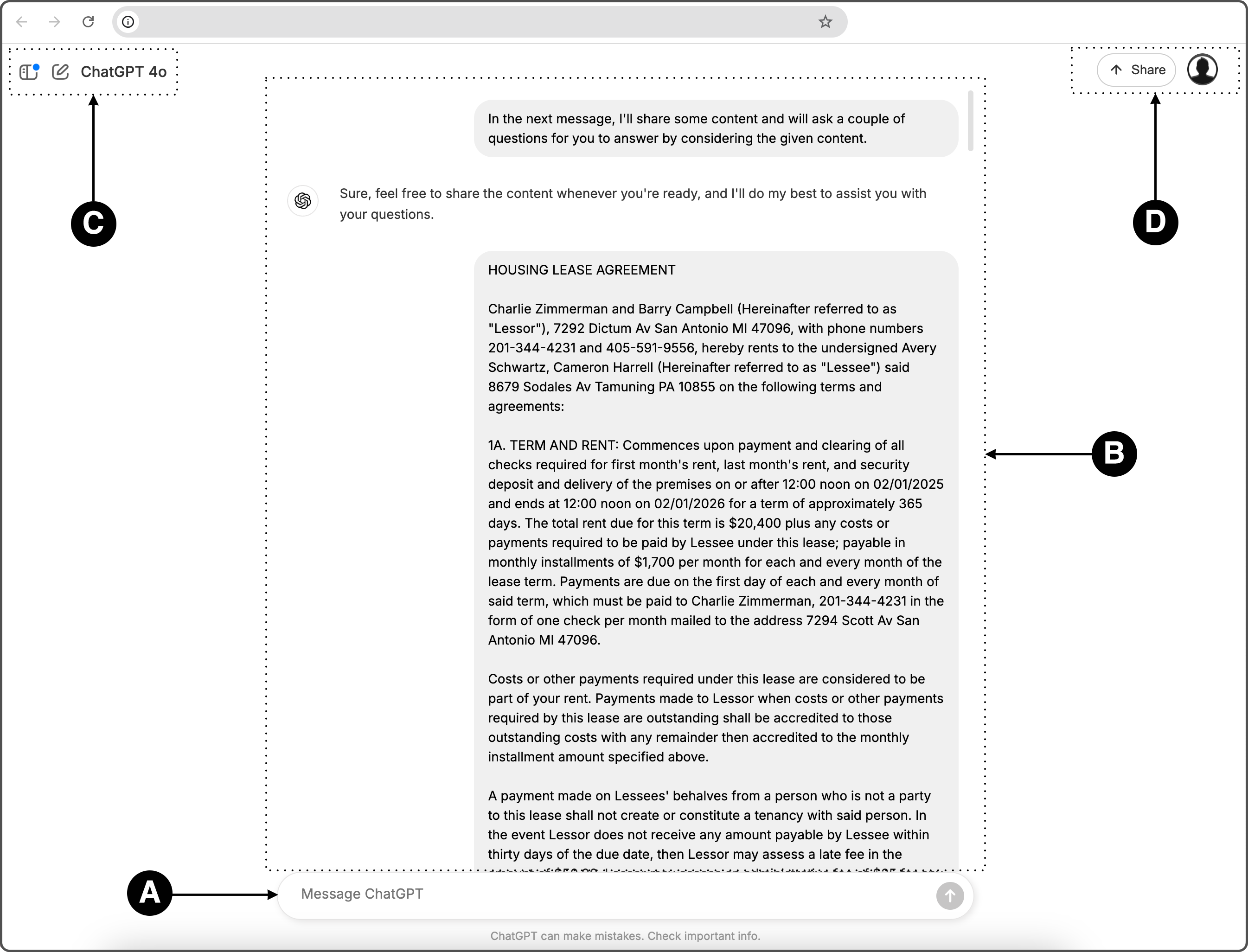}
  \caption{ChatGPT Interface Simulation. The interface includes (A) a message input box, (B) a conversation panel, (C) static icons for sidebar, new chat, and model name, and (D) a static chat-share button and an interactive profile icon.}
  \label{fig:without_panel}
  \Description{Screenshot of the simulated ChatGPT interface. The interface includes: (A) a message input box at the bottom, (B) a central conversation panel displaying user messages and chatbot responses, (C) static icons for sidebar, new chat, and model name at the top left, and (D) a static chat-share button and interactive profile icon at the top right.}
\end{figure}

\begin{figure}[h]
  \centering
  \includegraphics[width=\linewidth]{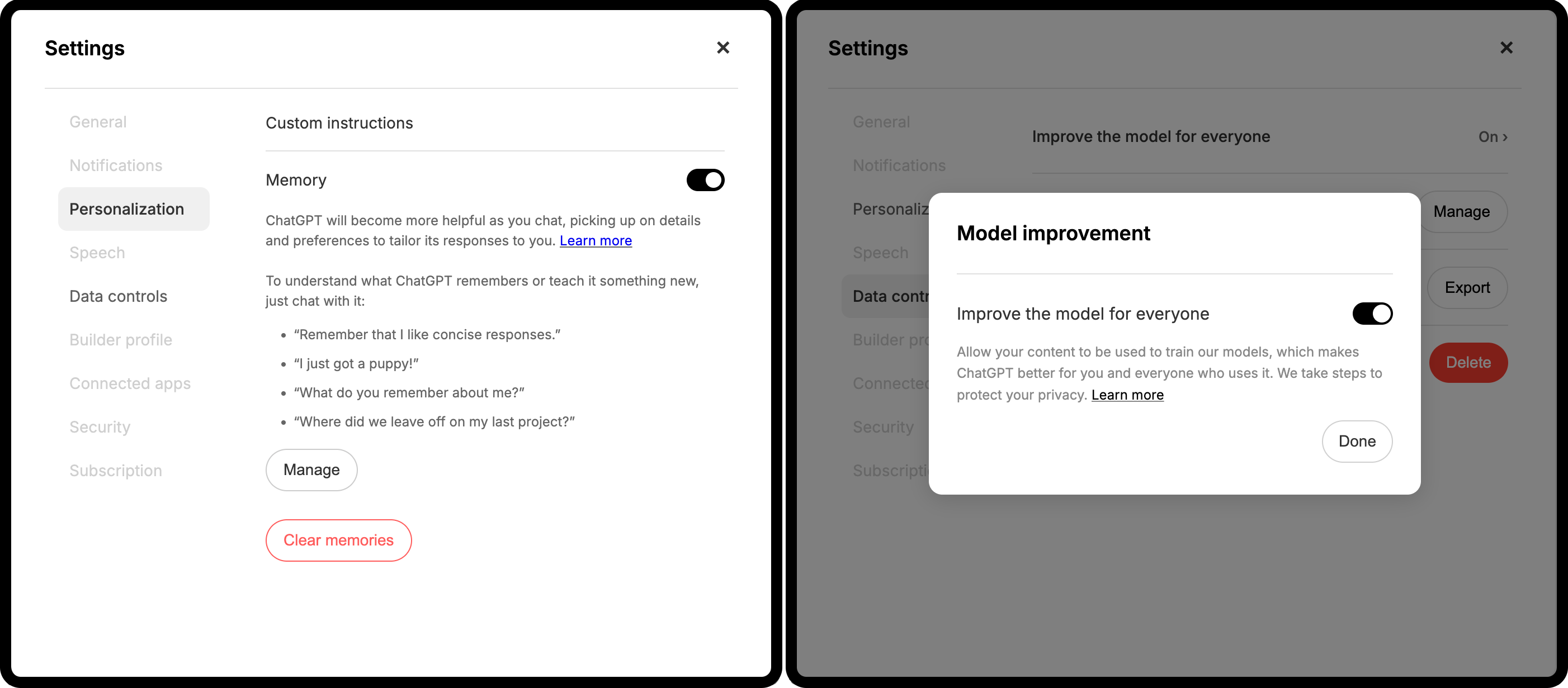}
  \caption{Simulated ChatGPT settings panels for built-in privacy controls: (left) Personalization panel for enabling or disabling memory, with a functional toggle and static buttons; (right) Data Control panel for opting in or out of content sharing for model training, with an interactive toggle.}
  \label{fig:settings-panel}
  \Description{Two screenshots of simulated ChatGPT settings panels for built-in privacy controls. The first shows the ‘Personalization’ tab for enabling or disabling memory. From top to bottom, it includes: a heading labeled ‘Custom Instructions’; a functional toggle button labeled ‘Memory’; a description reading ‘ChatGPT will become more helpful as you chat, picking up on details and preferences to tailor its responses to you. Learn More (hyperlink). To understand what ChatGPT remembers or teach it something new, just chat with it’; and two static buttons labeled ‘Manage’ and ‘Clear memories’. The second screenshot shows the ‘Data Control’ tab for opting in or out of content sharing for model training. It displays a pop-up panel with: a heading labeled ‘Model Improvement’; an interactive toggle button labeled ‘Improve the model for everyone’; and a description reading ‘Allow your content to be used to train our models, which makes ChatGPT better for you and everyone who uses it. We take steps to protect your privacy. Learn More (hyperlink).’}
\end{figure}

Figure \ref{fig:without_panel} shows our simulation of \textit{ChatGPT}’s interface without the privacy notice panel, which was used in the first study session. 
Figure \ref{fig:settings-panel} shows the simulated settings panel for built-in privacy controls: (left) disabling memory and (right) opting out of sharing content for model training. These panels were available in both study sessions.

\subsection{User Task Details}
\label{appendix_user_tasks}

Table \ref{tab:task-overview} shows detailed description of user tasks and the embedded sensitive information including their types, frequencies, and subjects---i.e., who the information pertains or belongs to.

\setlength{\tabcolsep}{6.5pt}
\begin{table*}
  \caption{User Tasks. Columns (L$\rightarrow$R): task assignment version (V), task description, types of sensitive info contained in each task (Sensitive Info), frequency---of occurrences---of each type (F), and Info Subjects (who the info pertains/belongs to).}
  \label{tab:task-overview}
  \begin{tabular}{p{0.2cm}p{7.8cm}p{2.2cm}p{0.3cm}p{2.2cm}}
    \toprule
    \textbf{V} & \textbf{Task Description} & \textbf{Sensitive Info} & \textbf{F} & \textbf{Info Subject} \\
    \toprule
    A 
    & 
    \textbf{Task 1:} Email Analysis
    \begin{itemize}[label=\textbullet,leftmargin=*,itemsep=0pt,topsep=0pt,parsep=0pt,partopsep=0pt]
        \item \textbf{Step 1:} Classifying seven emails\footnotemark[1] by their sentiments
        \item \textbf{Step 2:} Summarizing two separate email threads
    \end{itemize}
    & 
    Name\newline
    Email address\newline
    Phone number\newline
    Physical address
    &
    61\newline
    23\newline
    6\newline
    4
    & 
    Others (e.g., employees)
    \\
    \midrule
    A 
    & 
    \textbf{Task 2:} Searching Housing Documents
    \begin{itemize}[label=\textbullet, leftmargin=*,itemsep=0pt,topsep=0pt,parsep=0pt,partopsep=0pt]
        \item \textbf{Step 1:} Finding party restrictions from their housing contract (9-page doc)
        \item \textbf{Step 2:} Finding maintenance information from the same housing contract and a welcome letter (5-page doc)
    \end{itemize}
    & 
    Name\newline
    Email address\newline
    Phone number\newline
    Physical address\newline
    SSN\newline
    Date of birth
    & 
    38\newline
    9\newline
    16\newline
    16\newline
    4\newline
    4
    &
    Self, Others (e.g., friend, home owner)
    \\
    \midrule
    A 
    & 
    \textbf{Task 3:} Planning a Trip
    \begin{itemize}[label=\textbullet, leftmargin=*,itemsep=0pt,topsep=0pt,parsep=0pt,partopsep=0pt]
        \item \textbf{Step 1:} Finding timing conflicts between their travel itinerary (2-page doc) and semester schedule (1-page doc)
        \item \textbf{Step 2:} Drafting an email to resolve the conflicts and provide their contact details for reservation purposes
    \end{itemize}
    & 
    Name\newline
    Email address\newline
    Phone number\newline
    Physical address\newline
    Date of birth
    & 
    5\newline
    1\newline
    3\newline
    7\newline
    2
    &
    Self, Others (e.g., friend)
    \\
    \midrule
    B 
    & 
    \textbf{Task 1:} Customer Complaint Analysis
    \begin{itemize}[label=\textbullet, leftmargin=*,itemsep=0pt,topsep=0pt,parsep=0pt,partopsep=0pt]
        \item \textbf{Step 1:} Classifying seven customer complaints\footnotemark[2] by relevant departments
        \item \textbf{Step 2:} Summarizing recommendations for the company from two separate groups of customer complaints
    \end{itemize}
    & 
    Name\newline
    Email address\newline
    Phone number\newline
    Physical address\newline
    Date of birth
    & 
    18\newline
    6\newline
    4\newline
    4\newline
    2
    &
    Others (e.g., customers)
    \\
    \midrule
    B 
    & 
    \textbf{Task 2:} Searching Car \& Housing Documents
    \begin{itemize}[label=\textbullet, leftmargin=*,itemsep=0pt,topsep=0pt,parsep=0pt,partopsep=0pt]
        \item \textbf{Step 1:} Finding costs and fees related to purchasing a leased car from their car lease contract (15-page doc)
        \item \textbf{Step 2:} Finding information about utilities and responsibilities from their housing contract (9-page doc)
    \end{itemize}
    & 
    Name\newline
    Email address\newline
    Phone number\newline
    Physical address\newline
    SSN\newline
    Date of birth
    & 
    23\newline
    4\newline
    10\newline
    12\newline
    4\newline
    4
    &
    Self, Others (e.g., friend, home owner)
    \\
    \midrule
    B 
    & 
    \textbf{Task 3:} Planning Teeth Removal
    \begin{itemize}[label=\textbullet, leftmargin=*,itemsep=0pt,topsep=0pt,parsep=0pt,partopsep=0pt]
        \item \textbf{Step 1:} Finding out-of-pocket expenses from their insurance document (6 pages) and an email from their doctor.
        \item \textbf{Step 2:} Drafting an email to their doctor providing insurance details and asking if the insurance is accepted
    \end{itemize}
    & 
    Name\newline
    Email address\newline
    Phone number\newline
    Physical address\newline
    Date of birth
    & 
    8\newline
    2\newline
    2\newline
    1\newline
    5
    &
    Self, Others (e.g., doctor)
    \\
    \bottomrule
  \end{tabular}
\end{table*}
\footnotetext[1]{Drawn from the \href{http://www.enron-mail.com/email/}{\emph{Enron Email Corpus}} and modified as needed.}
\footnotetext[2]{Drawn from a \href{https://www.kaggle.com/datasets/venkatasubramanian/automatic-ticket-classification}{\emph{Ticket Classification Dataset}} and modified as needed.}

\section{Supplementary Discussion Materials}
\label{appendix_discussion}
\begin{figure}[h]
  \centering
  \includegraphics[width=0.8\linewidth]{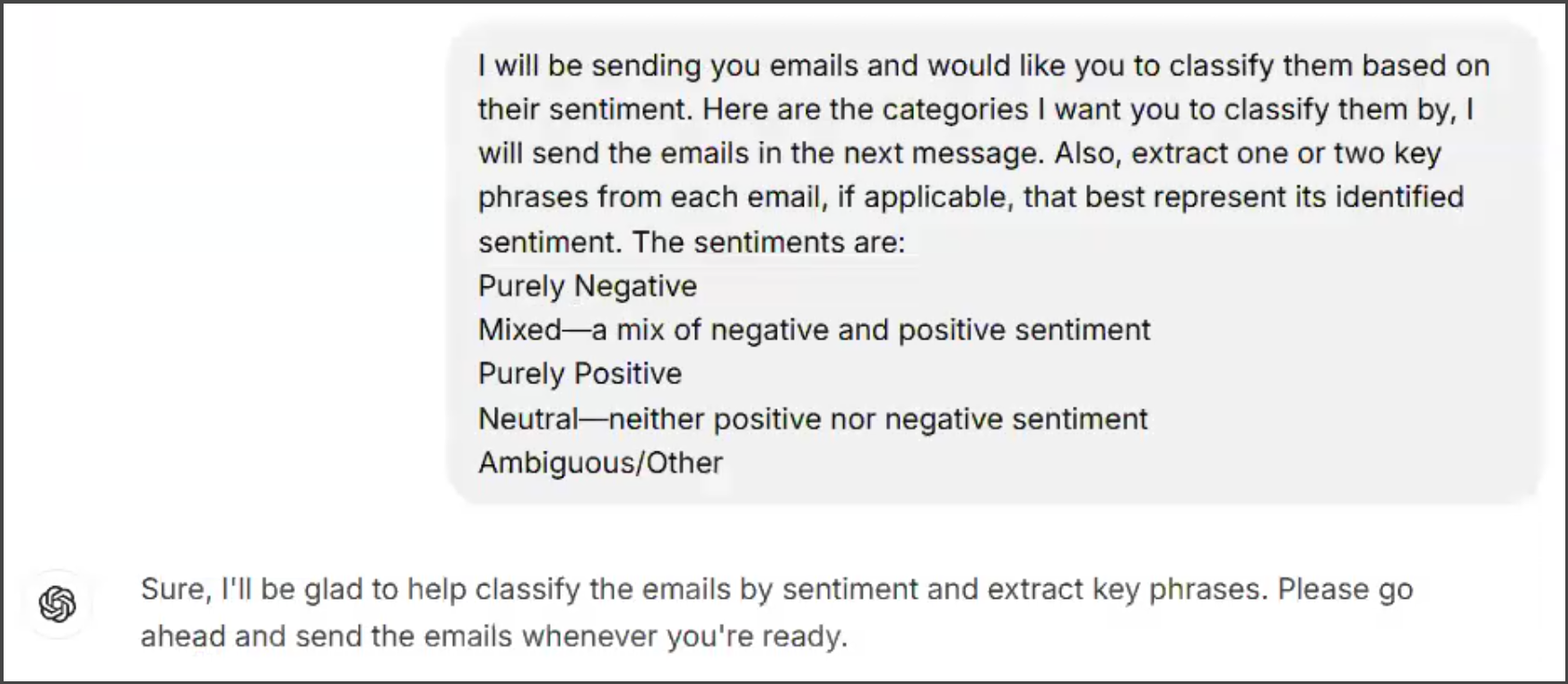}
  \caption{Example of incremental prompting from \textit{P6}’s chatbot interaction in the email classification task (Assignment A, Task 1), where the chatbot can prompt the user to withhold sender details in subsequent messages instead of leaving the request open-ended.}
  \label{fig:incremental_prompting}
  \Description{Example of incremental prompting from P6's chatbot interactions in the email classification task (Assignment A, Task 1). The user submitted the message: ‘I will be sending you emails and would like you to classify them based on their sentiment. Here are the categories I want you to classify them by, I will send the emails in the next message. Also, extract one or two key phrases from each email, if applicable, that best represent its identified sentiment. The sentiments are: Purely Negative, Mixed—a mix of negative and positive sentiment, Purely Positive, Neutral—neither positive nor negative sentiment, and Ambiguous/Other’ and the chatbot responded with ‘Sure, I'll be glad to help classify the emails by sentiment and extract key phrases. Please go ahead and send the emails whenever you're ready.’}
\end{figure}

\end{document}